\documentstyle[preprint,aps,eqsecnum,12pt]{revtex}

\author{F. T. Brandt and J. Frenkel\\
Instituto de F\'\i sica, Universidade de S\~ao Paulo,\\
S\~ao Paulo, 01498 SP, Brasil}
\title{The graviton self-energy in thermal quantum gravity
}

\begin{document}

\maketitle
\begin{abstract}
We show generally that in thermal gravity, the one-particle irreducible
2-point function depends on the choice of the basic graviton fields. We
derive the relevant properties of a physical graviton self-energy, which is
independent of the parametrization of the graviton field. An explicit
expression for the graviton self-energy at high-temperature is given to
one-loop order.
\end{abstract}

\section{Introduction}

\label{intro}

The high-temperature properties of quantum gravity are of some interest,
both in their own right as well as for their potential cosmological
applications. If the temperature $T$ is well below the Planck scale,
perturbation theory can be used to calculate the n-point thermal graviton
functions, with internal lines which correspond to matter and fields in
thermal equilibrium. These Green functions have been studied previously \cite
{GroPerGribDon,KikuMorTsu,Rebhan,FrenTay}, and show a leading $T^4$ behavior.

One of the most noteworthy aspects in field theory is the freedom of choice
in the fundamental fields. In the case of quantum gravity this freedom is
extensive, so that one may use as basic fields the metric tensor $g_{\mu \nu
}$, or its inverse $g^{\mu \nu }$, or any other function of $g_{\mu \nu }$.
{}From the point of view of the S-matrix elements at zero temperature, such
different choices lead to the same result \cite{HoVel}. In thermal field
theories at high-temperatures, the leading contributions to the one-particle
irreducible Green functions are gauge-independent quantities, describing
the physical properties of hot plasma \cite{WeldKapu,BraaPis,FrenTay1}. In
thermal quantum gravity, these contributions are also gauge invariant \cite
{Rebhan,FrenTay}, being however dependent on the parametrization of the
graviton fields. For instance, as pointed out by Kikuchi, Moriya and
Tsukahara \cite{KikuMorTsu}, the time components of the graviton two-point
function $\Pi ^{00,00}$ evaluated to one-loop order
at zero momenta, depend on the choice of
the basic graviton fields. Nevertheless, these authors have found that in
the high-temperature limit, the quantity $\Pi ^{00,00}-\Pi ^\rho {}_{\rho
,}{}^{00}$ is independent of these choices.

The main purpose of this work is to generalize the above features, for all
components of the 1PI 2-point function $\Pi ^{\mu \nu,\;\alpha \beta }$
evaluated at arbitrary momenta $k$. As we will argue, the
parametrization-dependence of $\Pi^{\mu \nu,\;\alpha \beta }$ is related to
the non-vanishing of the 1-point function (tadpole) in thermal quantum
gravity. In section \ref{sec2} we present the results of explicit one-loop
calculations, showing that the graviton self-energy given by:
\begin{equation}
\label{PiBar}\bar \Pi ^{\mu \nu ,\alpha \beta }\left( k\right) \equiv \Pi
^{\mu \nu ,\alpha \beta }\left( k\right) -\frac 14\left( \eta ^{\mu \alpha
}\Pi ^\rho {}_{\rho ,}{}^{\nu \beta }+\eta ^{\mu \beta }\Pi ^\rho {}_{\rho
,}{}^{\nu \alpha }+\eta ^{\nu \alpha }\Pi ^\rho {}_{\rho ,}{}^{\mu \beta
}+\eta ^{\nu \beta }\Pi ^\rho {}_{\rho ,}{}^{\mu \alpha }\right) ,
\end{equation}
yields in the most common cases a quantity which does not depend on the
choices of the basic graviton fields. In section \ref{sec3}, we discuss the
general conditions under which the expression (\ref{PiBar}) represents a
physical amplitude, which is independent of the graviton parametrization.
These are characterized by the fact that the graviton self-energy is
described by a traceless function, which is transverse with respect to the
external momenta at $k^2=0$.

\section{The graviton self-energy}

\label{sec2}

In order to derive the expression of the graviton self-energy to one-loop
order, we consider first the Feynman diagrams contributing to $\Pi $, which
are shown in Fig.\ref{fig1}. Here, the external lines denote the
gravitational field and the internal lines represent hot matter and
gravitational fields in thermal equilibrium.

There exists in the literature several methods for the evaluation of the
leading contributions in the high temperature limit, i.e. when $k_0,$ $%
\left| \vec k\right| \ll T$. In this domain all masses can be neglected, so
one can consider the particles as being effectively massless. The method we
use is that of reference \cite{BranFren}, in which the Green functions are
related to the forward scattering amplitudes of the thermal gravitons, with
on-shell momenta $q$. This method simplifies considerably the calculations
in the present case. Here we will just state the main results.

The relevant Feynman rules follow from the Einstein Lagrangian:
\begin{equation}
\label{Einstein}{\cal L}_{grav}=\frac 1{2\kappa ^2}\sqrt{-g}R,
\end{equation}
where $\kappa =\sqrt{32\pi G}$. Since the leading thermal contributions are
gauge independent \cite{FrenTay}, it is convenient for computational
purposes to fix the gauge by choosing:
\begin{equation}
\label{Lfix}{\cal L}_{fix}=-\frac 1{\kappa ^2}\left( \partial _\mu \sqrt{-g}%
g^{\mu \nu }\right) ^2.
\end{equation}
Then, subjecting ${\cal L}_{fix}$ to an infinitesimal gauge transformation
characterized by the parameter $\xi $, one obtains the Faddeev-Popov ghost
Lagrangian:
\begin{equation}
\label{Ghost}{\cal L}_{ghost}=\left( \partial _\mu \bar \chi _\nu \right)
\frac{\partial \left( \sqrt{-g}g^{\mu \nu }\right) }{\partial \xi ^\lambda }%
\chi _\lambda .
\end{equation}

We will show that the expression for $\Pi $ depends in general on the
representation of the graviton fields. To this end, let us define the basic
graviton field $h$ by \cite{Rebhan}:
\begin{equation}
\label{gh}g_{\mu \nu }=\eta _{\mu \nu }+\kappa h_{\mu \nu }.
\end{equation}
With the help of the results derived in reference \cite{BranFren}, we can
express the corresponding graviton 2-point function in the form:
\begin{equation}
\label{Pih}
\begin{array}{llll}
\Pi _h^{\mu \nu ,\alpha \beta }\left( k\right)  & = & \displaystyle\omega
\frac{\pi ^2T^4}{120}\kappa ^2\int \frac{d\Omega }{4\pi } & \displaystyle%
\left( \frac{k^\alpha Q^\mu Q^\nu Q^\beta }{k\cdot Q}+\frac{k^\beta Q^\mu
Q^\nu Q^\alpha }{k\cdot Q}\right.  \\  &  &  & \displaystyle+\frac{k^\mu
Q^\nu Q^\alpha Q^\beta }{k\cdot Q}+\frac{k^\nu Q^\mu Q^\alpha Q^\beta }{%
k\cdot Q}-\frac{k^2Q^\mu Q^\nu Q^\alpha Q^\beta }{\left( k\cdot Q\right) ^2}
\\  &  &  & \displaystyle-\left. \eta ^{\nu \beta }Q^\mu Q^\alpha -\eta
^{\nu \alpha }Q^\mu Q^\beta -\eta ^{\mu \beta }Q^\nu Q^\alpha -\eta ^{\mu
\alpha }Q^\nu Q^\beta \right) ,
\end{array}
\end{equation}
where $\omega $ is a weight factor given by the total number of degrees of
freedom of the thermal particles, being 2 for a graviton field. $Q^\alpha
=q^\alpha /\left| \vec q\right| =\left( 1,\hat Q\right) $ and ${\textstyle %
\int d\Omega }$ denotes angular integration over the 3-dimensional unit
vector $\hat Q$.

On the other hand, we may choose the basic graviton field $\tilde h$ defined
by \cite{CapperLeibbrandt}: $\sqrt{-g}\,g^{\mu \nu }=\eta ^{\mu \nu }+\kappa
\,\tilde h^{\mu \nu }$. Using Eq. (\ref{gh}), we can express $\tilde h$ in
terms of $h$ as follows:
\begin{equation}
\label{hTilh}\tilde h^{\mu \nu }=-h^{\mu \nu }+ \frac
12h_{\;\;\lambda}^\lambda \,\eta ^{\mu \nu }- \frac
12h_{\;\;\lambda}^\lambda \,h^{\mu \nu }+ h^{\mu\lambda} \,
h_{\;\;\lambda}^{\nu}+ \frac 18\left(h_{\;\;\lambda}^\lambda\right)^2 \,\eta
^{\mu \nu } - \frac 14h_{\alpha\beta} h^{\alpha\beta}\,\eta ^{\mu \nu
}+\cdots\;\; .
\end{equation}
We then obtain for the corresponding graviton 2-point function the result:

\begin{equation}
\label{Pihtil}\Pi _{\tilde h}^{\mu \nu ,\alpha \beta }\left( k\right) =\Pi
_h^{\mu \nu ,\alpha \beta }\left( k\right) +\omega \frac{\pi ^2T^4}{120}%
\kappa ^2\int \frac{d\Omega }{4\pi }\left( \eta ^{\nu \beta }Q^\mu Q^\alpha
+\eta ^{\nu \alpha }Q^\mu Q^\beta +\eta ^{\mu \beta }Q^\nu Q^\alpha +\eta
^{\mu \alpha }Q^\nu Q^\beta \right) ,
\end{equation}
which is clearly different from the one given in (\ref{Pih}). A similar
conclusion is obtained by using as the basic graviton field $h^{*}$, defined
by the inverse metric tensor $g^{\mu \nu }=\eta ^{\mu \nu }+\kappa
\,h^{*\;\mu \nu }$. From Eq. (\ref{gh}), we see that $h^{*}$ can be
expressed in a power series of $h$ given by:
\begin{equation}
\label{hStarh}h^{*\mu \nu }=-h^{\mu \nu }+h^{\mu \alpha }\,h_{\;\;\alpha
}^\nu +\cdots \;\;.
\end{equation}
However, the graviton self-energy defined by equation (\ref{PiBar}) is found
to be independent of these choices. In the above cases, $\bar \Pi $ is given
by one and the same expression, namely:
\begin{equation}
\label{PiBar2}
\begin{array}{llll}
\bar \Pi ^{\mu \nu ,\alpha \beta }\left( k\right)  & = & \displaystyle\omega
\frac{\pi ^2T^4}{120}\kappa ^2\int \frac{d\Omega }{4\pi } & \displaystyle%
\left[ \frac{k^\mu Q^\nu Q^\alpha Q^\beta }{k\cdot Q}+\frac{k^\nu Q^\mu
Q^\alpha Q^\beta }{k\cdot Q}\right.  \\  &  &  & \displaystyle\frac{k^\alpha
Q^\mu Q^\nu Q^\beta }{k\cdot Q}+\frac{k^\beta Q^\mu Q^\nu Q^\alpha }{k\cdot Q%
}-\frac{k^2Q^\mu Q^\nu Q^\alpha Q^\beta }{\left( k\cdot Q\right) ^2} \\  &
&  & \displaystyle-\left. \frac 12\left( \eta ^{\nu \beta }Q^\mu Q^\alpha
+\eta ^{\nu \alpha }Q^\mu Q^\beta +\eta ^{\mu \beta }Q^\nu Q^\alpha +\eta
^{\mu \alpha }Q^\nu Q^\beta \right) \right]
\end{array}
\end{equation}
The angular integrations can be performed with the help of the formulas
given in Appendix C of reference\cite{BranFren}. Then, the result for the
graviton self-energy $\bar \Pi $ can be expressed in terms of a basis of 14
independent tensors $T_i^{\mu \nu ,\,\alpha \beta }$, built from $\eta ^{\mu
\nu }$, $u^\mu \equiv \delta _0^\mu $ and:
\begin{equation}
\label{BigK}K^\mu \equiv \left( \frac{k_0}{\left| \vec k\right| },\,\hat
k\right) \equiv \left( r,\,\hat k\right) .
\end{equation}
Their explicit forms, presented in Tab. \ref{tab1}, were previously given by
Rebhan in Ref \cite{Rebhan}. In the basis of tensors listed in Tab. \ref
{tab1}, the expression for the graviton self-energy can be written as
follows:
\begin{equation}
\label{DecPi}\bar \Pi ^{\mu \nu ,\alpha \beta }\left( k\right) =\omega \frac{%
\pi ^2T^4}{30}\kappa ^2\sum_{i=1}^{14}\bar c_i\left( K,r\right) T_i^{\mu \nu
,\,\alpha \beta },
\end{equation}
where the 14 structure functions $\bar c_i\left( K,r\right) $ are given by:
\begin{mathletters}
\begin{eqnarray}
\label{str1}
\bar c\,_1&=&{\frac 1{12}}-
{\frac{{K^2}}{{24}}}+
{\frac{{{K^4}}}8L\left( r\right) } \\
\label{str2}\bar c\,_2&=&-{\frac16}+
{\frac{{K^2}}{{12}}}-
{\frac{{5\,K^4}}{{24}}}+
{\frac{{5\,{K^6}}}8L\left( r\right) } \\
\label{str3}\bar c\,_3&=&-{\frac{{{K^2}}}3}+
{\frac{{7\,{K^4}}}{{12}}}-
{\frac{{35\,{K^6}}}{{24}}}+
{\frac{{35\,{K^8}}}8L\left( r\right) } \\
\label{str4}\bar c\,_4&=&-
{\frac{{{K^2}}}{{24}}}+
{\frac{{{K^4}}}8L\left( r\right) } \\
\label{str5}\bar c\,_5&=&{\frac{{K^2}}{{12}}}-
{\frac{{5\,{K^4}}}{{24}}}+
{\frac{{5\,{K^6}}}8L\left( r\right) } \\
\label{str6}\bar c\,_6&=&\left( -{\frac1{{12}}}+
{\frac{{5\,{K^2}}}{{24}}}-
{\frac{{5\,{K^4}}}8L\left( r\right) }\right) \,r \\
\label{str7}\bar c\,_7&=&\left({\frac 13}-
{\frac{{7\,{K^2}}}{{12}}}+
{\frac{{35\,{K^4}}}{{24}}}-
{\frac{{35\,{K^6}}}8L\left( r\right) }\right) \,r  \\
\label{str8}\bar c\,_8&=&-{\frac 1{{12}}}-
{\frac{{5\,{K^2}}}{{24}}}+
{\frac{{{K^2}}}2L\left( r\right) }+
{\frac{{5\,{K^4}}}8L\left(r\right) } \\
\label{str9}\bar c\,_9&=&{\frac 16}-
{\frac{{2\,{K^2}}}3}-
{\frac{{35\,{K^4}}}{{24}}}+
{\frac{{15\,{K^4}}}4L\left( r\right)}+
{\frac{{35\,{K^6}}}8L\left( r\right) } \\
\label{str10}\bar c\,_{10}&=&{\frac 16}-
{\frac{{2\,{K^2}}}3}-
{\frac{{35\,{K^4}}}{{24}}}+
{\frac{{15\,{K^4}}}4L\left( r\right)}+
{\frac{{35\,{K^6}}}8L\left( r\right) } \\
\label{str11}\bar c\,_{11}&=&\left( {\frac14}+
{\frac{{35\,{K^2}}}{{24}}}-
{\frac{{5\,{K^2}}}2L\left( r\right) }-
{\frac{{35\,{K^4}}}8L\left( r\right) }\right) \,r \\
\label{str12}\bar c\,_{12}&=&-{\frac{13}{{12}}}-
{\frac{{35\,{K^2}}}{{24}}}+
L\left( r\right)+
5\,{K^2\,L\left( r\right) }+
{\frac{{35\,{K^4}}}8L\left(r\right) } \\
\label{str13}\bar c\,_{13}&=&-{\frac 1{{12}}}-
{\frac{{5\,{K^2}}}{{24}}}+
{\frac{{{K^2}}}2L\left( r\right) }+
{\frac{{5\,{K^4}}}8L\left(r\right) } \\
\label{str14}\bar c\,_{14}&=&\left(- {\frac1{{12}}}+
{\frac{{5\,{K^2}}}{{24}}}-
{\frac{{5\,{K^4}}}8L\left( r\right) }\right) \,r
\end{eqnarray}
\end{mathletters}and we have defined:
\begin{equation}
\label{LFun}{{L\left( r\right) =\frac r2\ln \left( \frac{r+1}{r-1}\right) -1.%
}}
\end{equation}

We now recall \cite{Rebhan,FrenTay}, that the trace of the graviton
two-point function is related in general to the graviton 1-point function $%
\Gamma ^{\mu \nu }$, which is traceless. For instance, in the representation
given by the $h$ field, the trace of the function $\Pi $ is connected to $%
\Gamma $ by the Ward identity:
\begin{equation}
\label{trPih}\Pi _{_h\,\rho ,}^\rho \,^{\mu \nu }\left( k\right) =-\kappa
\;\Gamma _{_h\,}^{\mu \nu }=-\kappa ^2\omega
\frac{\pi ^2T^4}{180}\left( 4\,\eta
^{\mu 0}\eta ^{\nu 0}-\eta ^{\mu \nu }\right) ,
\end{equation}
which is a consequence of the invariance under Weyl transformations. On the
other hand, it is easy to verify using (\ref{PiBar}) that the trace of the
graviton self-energy vanishes:
\begin{equation}
\label{trPiBar}\bar \Pi _{\;\;\rho ,}^\rho \,^{\mu \nu }\left( k\right) =0.
\end{equation}

This property indicates that the graviton self-energy is related to the
energy-momentum tensor corresponding to massless particles, which is also
traceless \cite{Rebhan,FrenTay}. Indeed, using equation (\ref{PiBar2}), we
obtain:
\begin{equation}
\label{Trans}k_\alpha k_\beta \bar \Pi ^{\mu \nu ,\alpha \beta }\left(
k\right) =-\frac{\kappa ^2}4k^2T_{\left( 0\right) }^{\mu \nu },
\end{equation}
where $T_{\left( 0\right) }^{\mu \nu }$ denotes the background
energy-momentum tensor:
\begin{equation}
\label{TenBack}T_{\left( 0\right) }^{\mu \nu }={\rm diagonal}\left( \rho ,%
{\rm \;p,\;p,\;p}\right)
\end{equation}
and $\rho =\omega \pi ^2T^4/30$ is the energy density of the thermal
particles, with the pressure assuming the maximal value $\rho /3$. It
follows from these relations that, at $k^2=0$, the graviton self-energy
satisfies the transversality condition,
\begin{equation}
\label{Tran1}\left. k_\alpha k_\beta \bar \Pi ^{\mu \nu ,\alpha \beta
}\left( k\right) \right| _{k^2=0}=0.
\end{equation}

As a consequence of invariance under general coordinate transformations, the
1PI graviton 2-point function $\Pi $ satisfies a Ward identity \cite
{Rebhan,FrenTay}, relating its divergence to the 1-point graviton function $%
\Gamma $. At zero temperature, it turns out that this function may
consistently be equated to zero \cite{CapperLeibbrandt}, so that an equation
like (\ref{Tran1}) holds directly for $\Pi $. But at finite temperatures,
the 2-point graviton function is not transverse since the tadpole $\Gamma $
is non-vanishing, being connected to the trace of $\Pi $ (cf. Eq. (\ref
{trPih})). In such cases, these Ward identities enforce the graviton
self-energy $\bar \Pi $ to obey the above physical transversality condition.

The properties expressed by equations (\ref{trPiBar}) and (\ref{Tran1}) are
important for the study of the relevant conditions which ensure the
invariance of the self-energy (\ref{PiBar}) under re-parametrizations of the
graviton fields.

\section{Discussion}

\label{sec3}

In order to investigate this problem in a general way, we consider the
effective {action~\cite{FrenTay}}:
\begin{equation}
\label{Seff}S_{eff}=\Gamma _{\prime }^{\alpha \beta }h_{_{\prime \,}\alpha
\beta }\left( 0\right) +\frac 12\int d^4k\Pi _{\prime }^{\mu \nu ,\,\alpha
\beta }\left( k\right) \,h_{\prime \,\mu \nu }\left( k\right) \,h_{\prime
\,\alpha \beta }\left( -k\right) +\cdots \;\;,
\end{equation}
which generates the one-particle irreducible Green functions of the field
theory at high temperature.
The most general re-parametrization of the graviton fields,
consistent with Lorentz invariance, can be written in a compact form as:
\begin{equation}
\label{repar}h_{\prime }^{\mu \nu }=a\,h^{\mu \nu }+b\,h_{\;\;\lambda
}^\lambda \,\eta ^{\mu \nu }+c\,h_{\;\;\lambda }^\lambda \,h^{\mu \nu
}+d\,h^{\mu \lambda }\,h_{\;\;\lambda }^\nu +e\,\left( h_{\;\;\lambda
}^\lambda \right) ^2\,\eta ^{\mu \nu }+f\,h_{\alpha \beta }h^{\alpha \beta
}\,\eta ^{\mu \nu }+\cdots \;\;,
\end{equation}
where $a$, $b$, $c$, $d$, $e$ and $f$ denote arbitrary constants and we have
chosen $h_{\mu \nu }$, for definiteness, to be the field defined by Eq. (\ref
{gh}). The examples given by Eqs. (\ref{hTilh}) and (\ref{hStarh}) represent
particular cases of this relation. Under the above transformation,
the action (\ref{Seff}) can be expanded in a power series of the field $h$.

On the other hand, the effective action should be invariant under a
re-parametrization of the graviton fields. Hence we must have that:
\begin{equation}
\label{Seff2}S_{eff}=\Gamma ^{\alpha \beta }\,h_{\alpha \beta }\left(
0\right) +\frac 12\int d^4k\ \Pi ^{\mu \nu ,\,\alpha \beta }\left(k\right)
\ h_{\mu \nu
}\left( k\right) \ h_{\alpha \beta }\left( -k\right) +\cdots \ ,
\end{equation}
where, for simplicity of notation, we drop the index $h$ from the above Green
functions. Identifying the corresponding terms
given by the forms (\ref{Seff}) and (\ref{Seff2}), we find:
\begin{equation}
\label{GGprime}\Gamma ^{\mu \nu }=a\,\Gamma _{\prime }^{\mu \nu },
\end{equation}
where we used the traceless property of $\Gamma $ [cf. Eq. (\ref{trPih})] and:
\begin{equation}
\label{PiPiprime}
\begin{array}{lll}
\Pi ^{\mu \nu ,\,\alpha \beta } \left(k\right)
& =a^2\Pi _{\prime }^{\mu \nu ,\,\alpha
\beta } \left(k\right)
& +a\,b\left( \Pi _{\prime }^{\mu \nu ,\ \rho }\,_\rho
\ \eta
^{\alpha \beta }+\Pi _{\prime }^{\alpha \beta ,\ \rho }\,_\rho
\ \eta ^{\mu\nu }\right)  \\
&  & +
\displaystyle c\left( \Gamma _{\prime }^{\mu \nu }\eta ^{\alpha \beta
}+\Gamma _{\prime }^{\alpha \beta }\eta ^{\mu \nu }\right)  \\  &  & +%
\displaystyle\frac d2\left( \Gamma _{\prime }^{\mu \alpha }\eta ^{\nu \beta
}+\Gamma _{\prime }^{\nu \beta }\eta ^{\mu \alpha }+\Gamma _{\prime }^{\nu
\alpha }\eta ^{\mu \beta }+\Gamma _{\prime }^{\mu \beta }\eta ^{\nu \alpha
}\right)
\end{array}
{}.
\end{equation}

{}From the above relations we obtain, with the aid of (\ref{trPih}), the Ward
identity obeyed by the function $\Pi _{\prime }$:
\begin{equation}
\label{trPiprime}\Pi_{\prime\;\rho ,\,}^{\rho\;\;\;\mu \nu }=-\frac{a+4c+2d}{%
a\left( a+4b\right) }\;\Gamma _{\prime }^{\mu \nu },
\end{equation}
where $a\neq 0$ and $a+4b\neq 0$, since we assume that (\ref{repar}) can be
inverted.

We now define $\bar \Pi _{\prime }$ analogously to equation (\ref{PiBar}),
with $\Pi $ replaced on the right hand side by $\Pi _{\prime }$. Then, with
the help of the Ward identity (\ref{trPiprime}), it is straightforward to
deduce the following relation:
\begin{equation}
\label{PiBarPiBarprime}
\begin{array}{lll}
\bar \Pi ^{\mu \nu ,\,\alpha \beta }\left( k\right) -a^2\bar \Pi _{\prime
}^{\mu \nu ,\,\alpha \beta }\left( k\right)  & =\displaystyle\left( c-\frac{%
a+4c+2d}{a+4b}\;b\right)  & \left( \eta ^{\mu \nu }\Gamma _{\prime }^{\alpha
\beta }+\eta ^{\alpha \beta }\Gamma _{\prime }^{\mu \nu }\right.  \\
&  & -\left. \eta ^{\mu \alpha }\Gamma _{\prime }^{\nu \beta }-\eta ^{\mu
\beta }\Gamma _{\prime }^{\nu \alpha }\right.  \\
&  & -\left. \eta ^{\nu \alpha }\Gamma _{\prime }^{\mu \beta }-\eta ^{\nu
\beta }\Gamma _{\prime }^{\mu \alpha }\right)
\end{array}
\end{equation}
It is easy to see [cf. Eq. (\ref{trPiBar})] that the
trace of the expression on the left-hand side of the above equation
vanishes. This is consistent with the form appearing on its right hand side,
because $\Gamma _{\prime }$ is a traceless function.

As pointed out following Eq. (\ref{Tran1}), we should require
the physical amplitude associated with a spin-2 massless graviton to
satisfy the transversality condition:
\begin{equation}
\label{Tran2}\left. k_\alpha k_\beta \bar \Pi _{\prime }^{\mu \nu ,\alpha
\beta }\left( k\right) \right| _{k^2=0}=0,
\end{equation}
which reflects the underlying gauge invariance of the effective action. The
above constraint then implies the vanishing of the expression on the
right-hand side of Eq. (\ref{PiBarPiBarprime}), yielding the relation:
\begin{equation}
\label{condit}a\,b-a\,c+2\,b\,d=0.
\end{equation}
This is explicitly verified for all the representations of the graviton
fields discussed in the previous section. Consequently, under the physical
constraint expressed by (\ref{Tran2}), the Eq. (\ref{PiBarPiBarprime})
reduces to:
\begin{equation}
\label{PiBarPiBarprime2}\bar \Pi ^{\mu \nu ,\,\alpha \beta }\left( k\right)
-a^2\bar \Pi _{\prime }^{\mu \nu ,\,\alpha \beta }\left( k\right) =0.
\end{equation}
We next rescale the fields $h_{\prime }^{\mu \nu }$ to: $h_{\prime \prime
}^{\mu \nu }=a^{-1}\;h_{\prime }^{\mu \nu }$ [cf. Eq. (\ref{repar})]. Making
use of the same method which lead to
Eqs. (\ref{GGprime}) and (\ref{PiPiprime}), we readily find:
\begin{equation}
\label{GG2prime}\Gamma _{\prime \prime }^{\mu \nu }=a\,\Gamma _{\prime
}^{\mu \nu }
\end{equation}
and
\begin{equation}
\label{PiG2prime}
\Pi _{\prime \prime }^{\mu \nu ,\,\alpha\beta }\left(k\right)
=a^2\Pi _{\prime }^{\mu \nu ,\,\alpha \beta }\left(k\right)
\end{equation}
Then, it follows from (\ref{PiBarPiBarprime2}) and (\ref{PiG2prime}) that
the graviton self-energy will be invariant under general
re-parametrizations, subject to the constraint (\ref{condit}), in the sense
that:
\begin{equation}
\label{PiInv}\bar \Pi ^{\mu \nu ,\,\alpha \beta }\left( k\right) =\bar \Pi
_{\prime \prime }^{\mu \nu ,\,\alpha \beta }\left( k\right) ,
\end{equation}
where we have defined $\bar \Pi _{\prime \prime \text{ }}$ similarly to
$\bar\Pi_{\prime }$. We emphasize that this result is general,
following in consequence of the invariance of the
effective action under coordinate and Weyl transformations.

In conclusion, we wish to comment on the mechanism which enforces the above
property of the thermal graviton self-energy. To this end, using the Ward
identity (\ref{trPiprime}) together with the relations (\ref{GGprime}),
(\ref{GG2prime}) and
(\ref{PiG2prime}), we obtain from Eq. (\ref{PiInv}) that:
\begin{equation}
\label{PiInv1}\Pi _{\prime \prime}^{\mu \nu ,\alpha \beta }\left(k\right)=
\Pi^{\mu \nu ,\alpha \beta }\left(k\right)+
\frac \kappa 2
\left(\displaystyle\frac{2b-2c-d}{a+4b}\right)
\left( \eta ^{\mu \alpha }\Gamma
^{\nu \beta }+\eta ^{\mu \beta }\Gamma ^{\nu \alpha }+\eta ^{\nu \alpha
}\Gamma ^{\mu \beta }+\eta ^{\nu \beta }\Gamma ^{\mu \alpha }\right) ,
\end{equation}
which shows generally that the 1PI graviton 2-point function is dependent on
the parametrization of the graviton fields. Clearly, this behavior occurs
because of the non-vanishing of the 1-point function.

Using the above relation and Eq. (\ref{PiBar}), we can write the expression
(\ref{PiInv}) in the from:
\begin{equation}
\label{PiBarPiTad}\bar \Pi^{\mu \nu ,\alpha \beta }
\left( k\right) =\Pi
_{\prime \prime }^{\mu \nu ,\alpha \beta }\left( k\right) +\Pi _{\prime
\prime \,tad}^{\mu \nu ,\alpha \beta },
\end{equation}
with $\Pi_{\prime \prime \,tad}^{\mu \nu ,\alpha \beta }$ given by:
\begin{equation}
\label{PiTad}\Pi _{\prime \prime \,tad}^{\mu \nu ,\alpha \beta }\equiv \frac
\kappa 4
\left(\displaystyle\frac{a+4c+2d}{a+4b}\right)
\left( \eta ^{\mu \alpha }\Gamma
^{\nu \beta }+\eta ^{\mu \beta }\Gamma ^{\nu \alpha }+\eta ^{\nu \alpha
}\Gamma ^{\mu \beta }+\eta ^{\nu \beta }\Gamma ^{\mu \alpha }\right) .
\end{equation}

We may represent $\Pi _{\prime \prime \,tad}$ diagrammatically as shown in
Fig. \ref{fig2}. It is important to distinguish these graphs from the usual
tadpole contributions to the graviton self-energy, which are very ambiguous,
being actually infinite. In our case, the contributions given by Eq. (\ref
{PiTad}) are well defined and finite. Furthermore, these are also gauge
independent, being the same for internal scalars, quarks or gravitons, up to
numerical factors which just count degrees of freedom. We see that
$\Pi_{\prime \prime \,tad}$ depends on the choice
of the basic graviton fields
in a way that ensures $\bar \Pi$ to
be independent of this parametrization.
Hence, in order to obtain a physical self-energy, one must consider in
addition to the 1PI 2-point function, also the corresponding tadpole
contributions. It is interesting to note that, when $a+4c+2d=0$, the graviton
self-energy can be identified with the proper 2-point function, since in
this case $\Pi _{\prime \prime\; tad}$ vanishes.

Finally, we mention that tadpole-free graviton fields might describe the
metric perturbations at high temperatures,
in a given background spacetime. Such fields arise naturally in the context
of a radiation dominated Robertson-Walker universe \cite{Rebhan,BirDav}.

\acknowledgements The authors are grateful to Professor J. C. Taylor for a
most valuable help throughout this work. They would also like to thank CNP$_{
\text{q}}$ (Brasil) for a grant.

\begin{table}
  \begin{tabular}{lcc}\hline
$T^{\mu\nu ,\, \alpha\beta}_1=\eta ^{\alpha \nu }\,\eta ^{\beta \mu }+
\eta ^{\alpha \mu }\,\eta ^{\beta \nu }$  \\
$T^{\mu\nu ,\, \alpha\beta}_2=u^\mu \,\left( u^\beta
\,\eta ^{\alpha \nu }+u^\alpha \,\eta ^{\beta \nu }\right) +
u^\nu \,\left( u^\beta \,\eta ^{\alpha \mu }+u^\alpha \,\eta ^{\beta \mu }
\right) $ \\
$T^{\mu\nu ,\, \alpha\beta}_3=u^\alpha \,u^\beta \,u^\mu \,u^\nu $ \\
$T^{\mu\nu ,\, \alpha\beta}_4=\eta ^{\alpha \beta }\,\eta ^{\mu \nu }$  \\
$T^{\mu\nu ,\, \alpha\beta}_5=u^\mu \,u^\nu \,\eta ^{\alpha \beta }+
u^\alpha \,u^\beta \,\eta ^{\mu \nu }$ \\
$T^{\mu\nu ,\, \alpha\beta}_6=u^\beta \,\left( K^\nu \,
\eta ^{\alpha \mu }+K^\mu \,\eta ^{\alpha \nu }\right) +
K^\beta \,\left( u^\nu \,\eta ^{\alpha \mu }+u^\mu \,\eta ^{\alpha \nu }
\right) $ \\
$\;\;\;\;\;\;\;\;\,\,\,\,\,\;+\;u^\alpha \,\left( K^\nu \,\eta ^{\beta \mu }+
K^\mu \,\eta ^{\beta \nu }\right) +K^\alpha \,\left( u^\nu
\,\eta ^{\beta \mu }+u^\mu \,\eta ^{\beta \nu }\right) $ \\
$T^{\mu\nu ,\, \alpha\beta}_7=K^\nu \,u^\alpha \,u^\beta \,u^\mu +
K^\mu \,u^\alpha \,u^\beta \,u^\nu +K^\beta \,u^\alpha \,u^\mu \,u^\nu +
K^\alpha \,u^\beta \,u^\mu \,u^\nu $ \\
$T^{\mu\nu ,\, \alpha\beta}_8=K^\beta \,K^\nu \,\eta ^{\alpha \mu }+
K^\beta \,K^\mu \,\eta ^{\alpha \nu }+
K^\alpha \,K^\nu \,\eta ^{\beta \mu }+
K^\alpha \,K^\mu \,\eta ^{\beta \nu }$ \\
$T^{\mu\nu ,\, \alpha\beta}_9=K^\mu \,K^\nu \,u^\alpha \,u^\beta +
K^\alpha \,K^\beta \,u^\mu \,u^\nu $ \\
$T^{\mu\nu ,\, \alpha\beta}_{10}=\left( K^\beta \,u^\alpha +
K^\alpha \,u^\beta \right) \,\left( K^\nu \,u^\mu +K^\mu \,u^\nu \right) $ \\
$T^{\mu\nu ,\, \alpha\beta}_{11}=K^\beta \,K^\mu \,K^\nu \,u^\alpha +
K^\alpha \,K^\mu \,K^\nu \,u^\beta +K^\alpha \,K^\beta \,K^\nu \,u^\mu +
K^\alpha \,K^\beta \,K^\mu \,u^\nu $ \\
$T^{\mu\nu ,\, \alpha\beta}_{12}=K^\alpha \,K^\beta \,K^\mu \,K^\nu $ \\
$T^{\mu\nu ,\, \alpha\beta}_{13}=K^\mu \,K^\nu \,\eta ^{\alpha \beta }+
K^\alpha \,K^\beta \,\eta ^{\mu \nu }$ \\
$T^{\mu\nu ,\, \alpha\beta}_{14}=\left( K^\nu \,u^\mu +K^\mu \,u^\nu
\right) \,\eta^{\alpha \beta }+\left( K^\beta \,u^\alpha +
K^\alpha \,u^\beta \right) \,\eta^{\mu \nu }$
   \end{tabular}
\vskip 0.7cm
\caption{A basis of 14 independent tensors\label{tab1}}
\end{table}

\begin{figure}

\caption{Lowest order contributions to the thermal 1PI graviton
2-point function. Wavy lines denote gravitons and dashed lines
represent ghost particles}\label{fig1}
\end{figure}

\begin{figure}
\caption{One-loop tadpole contributions of graviton and ghost
fields to $\Pi_{\prime\prime\; tad}$.
The black dot stands for terms proportional
to the $\eta$ functions in Eq. (3.16)}\label{fig2}
\end{figure}

\end{document}